# A Study of Bluetooth Access Control Based on NFT Soft Pairing

Zhiming Liang
*College of Electronics and Information Engineering*
*Shenzhen University*
Shenzhen, China
lzming0212@163.com

Bin Chen*
*College of Electronics and Information Engineering*
*Shenzhen University*
Shenzhen, China
bchen@szu.edu.cn
*Corresponding author

Ruijun Wu
*College of Electronics and Information Engineering*
*Shenzhen University*
Shenzhen, China
19892315202@163.com

Zhe Peng
*Department of Industrial and Systems Engineering*
*The Hong Kong Polytechnic University*
Hong Kong, China
jeffrey-zhe.peng@polyu.edu.hk

Chen Sun
*Wireless Network Research Department*
*Research and Development Center, Sony (China) Limited*
Beijing, China
chen.sun@sony.com

Shuo Wang
*Wireless Network Research Department*
*Research and Development Center, Sony (China) Limited*
Beijing, China
shuo.wang@sony.com

***Abstract*—This paper proposes a Non-Fungible Token (NFT) soft pairing framework for Bluetooth service access control. Unlike conventional Bluetooth systems where pairing implicitly grants persistent service access, the proposed approach decouples native Bluetooth pairing from authorization without modifying the underlying protocol stack. The framework introduces a three-layer architecture consisting of a Bluetooth layer for connectivity, a blockchain layer for trusted execution and on-chain state verification, and an application layer where NFT soft pairing defines the authorization logic. In this design, Non-Fungible Bluetooth Tokens (NFBTs) represent user-side access credentials, while Non-Fungible Device Tokens (NFDTs) represent device identities. Their bidirectional on-chain binding forms a revocable and verifiable NFT soft pairing relationship. During access, users prove ownership of valid NFBTs through challenge-response signatures, and devices verify the corresponding on-chain state before granting service access. A prototype implemented with MetaMask and Ethereum demonstrates secure authentication, dynamic revocation, acceptable latency, and gas-efficient credential issuance based on ERC1155.**

## I. Introduction

Bluetooth technology [1][2] has evolved from an early short-range data exchange protocol into a widely deployed service interface for audio devices, smart terminals, intelligent vehicles, and Internet of Things (IoT) systems. In many application scenarios, establishing a Bluetooth connection is no longer merely a communication process but also serves as a means of obtaining access to device services or resources. Therefore, Bluetooth access control has become an increasingly important security problem. Classical Bluetooth Basic Rate/Enhanced Data Rate (BR/EDR), which is the primary focus of this paper, is widely used in scenarios such as audio transmission and file sharing. For simplicity, it is hereafter referred to as Bluetooth.

The critical limitation of conventional Bluetooth security is not only that the pairing procedure is vulnerable to various attacks, including downgrade-to-Just Works (JW) attacks [3], method confusion attacks [4], pairing confusion attacks, and Stealtooth attacks [5], but also that access control is tightly coupled with pairing itself. Once a long-term Link Key is established [6], the peer device is often treated as authorized until the user manually removes the pairing record. As a result, a compromised or mistakenly trusted pairing relationship may be transformed into persistent service access.

In addition to vulnerabilities in the pairing procedure, conventional Bluetooth access control models are inherently static and coarse-grained. Once device pairing is completed, the long-term link key grants the peer device nearly unrestricted access privileges until manually revoked by the user. This process is cumbersome and highly prone to oversight in multi-device management scenarios.

Existing studies primarily focus on protocol-layer mitigations, which often suffer from poor compatibility, and some approaches target only specific attack scenarios, making them incapable of comprehensively defending against known attacks. Consequently, designing a universal communication protocol applicable to all wireless devices remains highly challenging, as the overall security of the system is constrained by the weakest device. In recent years, research efforts have shifted toward introducing additional authentication and access control mechanisms at the application layer, thereby enhancing overall system security without modifying the underlying protocols. Reference [7] proposes a certificate-based authentication mechanism. However, the approach relies on a centralized certificate authority, which introduces the risk of certificate forgery. Moreover, certificates suffer from issues such as complex lifecycle management, lack of transparent auditing, and difficulty in supporting dynamic fine-grained access control.

Blockchain technology, with its decentralized and tamper-resistant characteristics, provides a novel approach for implementing trustworthy access control [8]. Non-Fungible Tokens (NFTs), featuring unique on-chain identifiers and verifiable ownership properties, can serve as programmable digital access credentials for describing device access permissions and their lifecycle states [9]. Smart contracts enable the access management process to be transparent and automated, thereby reducing management and transaction costs [10].

To address this gap, we introduce NFT soft pairing as a new abstraction that decouples Bluetooth physical pairing from logical service authorization. In this model, native Bluetooth pairing is retained for connectivity, while access control is governed by an NFT-based soft pairing layer that establishes a bidirectional association between a user-side Non-Fungible Bluetooth Token (NFBT) and a device-side Non-Fungible Device Token (NFDT), forming a persistent and verifiable on-chain binding that determines service authorization. Based on this design, we develop an NFBT–NFDT soft pairing mechanism to support identity-based access control through blockchain-enforced token relationships. A complete prototype system is further implemented, including smart contracts, a MetaMask-integrated decentralized application (DApp), and a Bluetooth device authentication module, enabling challenge-response verification, dynamic access control, and lifecycle-based revocation.

## II. SYSTEM MODEL

As illustrated in Fig. 1, the proposed framework introduces a hierarchical three-layer architecture to decouple native Bluetooth pairing from service authorization. Layer 1, the Bluetooth Layer, handles pairing, connection, and communication to ensure protocol compatibility. Layer 2, the Blockchain Layer, serves as the decentralized trust anchor by executing smart contracts to manage authorization rules. Operating at the application layer, Layer 3, the Application Layer, serves as an external trusted authentication system. Within this infrastructure, NFBTs act as user credentials and NFDTs represent device identities. Native Bluetooth pairing serves only as a prerequisite for access and does not directly grant Bluetooth connection privileges. Only after a bidirectional NFT soft pairing relationship has been established can a user obtain Bluetooth connection authorization by proving ownership of the NFBT, thereby gaining access to the designated Bluetooth service. The proposed system consists of the following components:

Blockchain: A platform for deploying and executing smart contracts, ensuring secure recording and reliable storage of transaction information and user data, while simultaneously verifying access permissions.

NFBT Client: A DApp that provides users and Bluetooth service operators with functionalities for minting, purchasing, and connection establishment.

MetaMask: A Web3 gateway injected into the NFBT Client to manage accounts and private keys, enabling secure blockchain interaction through transaction confirmation, challenge signing, and authorization.

Users and User Devices: Users manage their devices and can acquire access permissions to Bluetooth services by purchasing NFBTs.

Bluetooth Service Operators: Responsible for deploying Bluetooth devices and smart contracts, maintaining a Bluetooth wallet, and minting NFBTs for Bluetooth devices.

Bluetooth Devices: Bluetooth devices provide Bluetooth services and verify users' access permissions through the blockchain, managing authorized users by querying blockchain data.

NFBT Contract: This contract comprises a minting module through which the operator batch-generates NFBTs based on device parameters, a sales module that enables users to purchase NFBTs using Ether (ETH), a verification module that performs on-chain validation of user access ownership for specific Bluetooth services, and an NFT soft pairing module that is responsible for requesting the linkage between NFBTs and NFDTs.

NFDT Contract: This contract includes a device registration module in which the operator binds the unique device identifier with an on-chain address to complete identity initialization, and an NFT soft pairing confirmation module that is used to confirm the linkage between NFDTs and NFBTs.

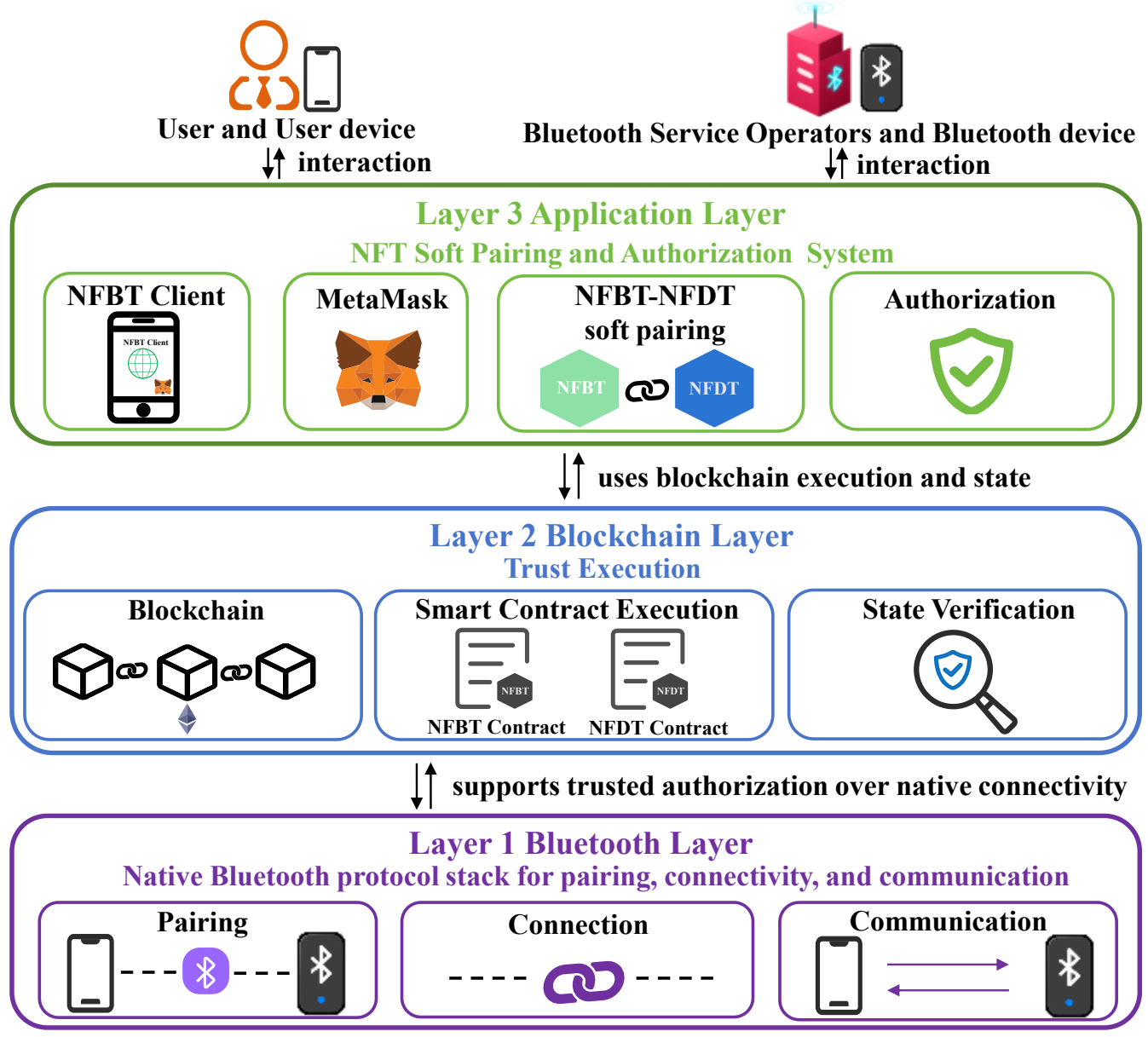

Fig. 1. Layered Architecture of the NFT Soft Pairing-based Bluetooth Access Control System.

As illustrated in Fig. 2, an on chain logical NFT soft pairing relationship must be established between the user's NFBT and the target device's NFDT prior to Bluetooth access verification. Once established, this relationship remains valid throughout the lifetime of the NFBT and serves as a prerequisite for subsequent access control verification without requiring repeated pairing. The user may then initiate an access request to the target Bluetooth service. The complete verification procedure is summarized as follows:

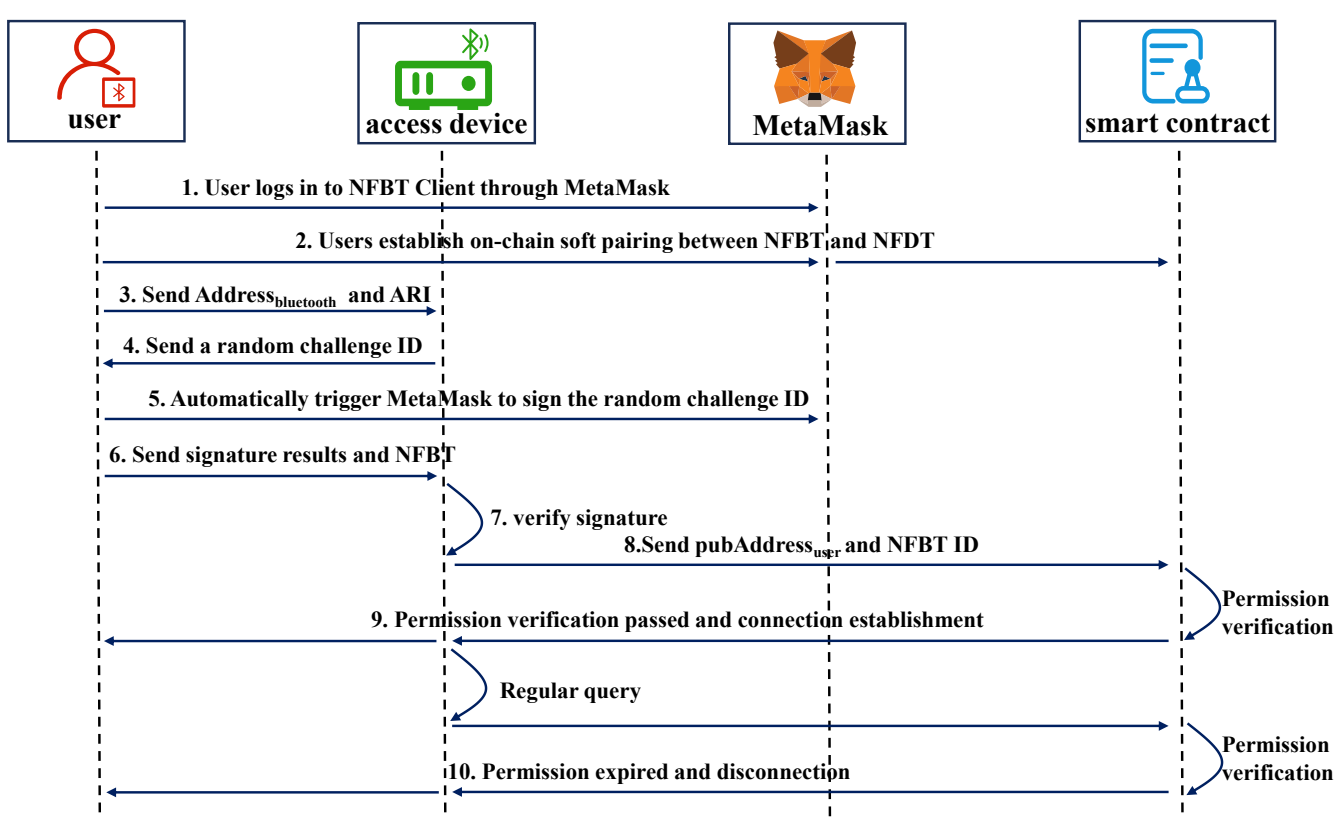

Fig. 2. Authentication and access control workflow in the NFT soft pairing framework.

1) The user logs into the NFBT Client via the MetaMask wallet, and the user interface displays all NFBTs currently owned by the user.

2) The user performs on-chain NFT soft pairing between the NFBT token ID and its corresponding device ID, which completes the bidirectional NFBT–NFDT soft pairing mechanism.

3) The user selects and clicks on the NFBTs corresponding to the Bluetooth device they wish to connect to, and sends the device's Bluetooth address along with an Access Request Identifier (ARI) to initiate the Bluetooth access permission verification process.

4) Upon receiving the ARI, the Bluetooth device locally generates a 256-bit random challenge ID and sends it to the user.

5) When the user device receives the random challenge ID, it automatically triggers MetaMask to sign the challenge ID.

6) After the user confirms the signing request in MetaMask, the user device immediately transmits the signature result along with the NFBT ID to the Bluetooth device for access permission verification, and waits for the verification outcome.

7) The Bluetooth device recovers the signer's address using the random challenge ID and the signature result according to the ECDSA algorithm.

8) The Bluetooth device queries the blockchain network using the signer's address, NFBT ID, and Bluetooth device identifier to verify the user's access permission.

9) Upon successful verification, the Bluetooth device establishes a Bluetooth connection with the user device and stores the user's Bluetooth address together with the remaining access duration in the active connection control set.

10) The Bluetooth device periodically queries the remaining access duration associated with authorized addresses in the control set. When the authorized service period expires, the system automatically terminates the connection with the corresponding address, revokes its Bluetooth service access, and removes the expired authorized address from the control set. Meanwhile, the NFT soft pairing is revoked through on-chain smart contract invocation, ensuring that the corresponding NFT soft pairing relationship is no longer valid during subsequent access verification procedures.

## III. SMART CONTRACT DESIGN

### *A. NFBT Contract*

This work adopts the ERC1155 standard [11] for minting NFBTs and NFDTs. As a multi-token management standard, ERC1155 enables the creation and management of both Fungible Tokens (FTs) and NFTs within a single smart contract. By supporting batch minting and transfer operations, ERC1155 significantly reduces on-chain gas consumption and alleviates network redundancy. These advantages render ERC1155 particularly suitable for tokenizing Bluetooth access permissions and enabling dynamic and revocable permission management.

**Algorithm 1:** mintNFBT

```
Input: deviceId, duration, location, price, quantity
1.    global tokenId
2.    require the device is registered and belongs to the caller
3.    if(isAuthorized(caller)) then
4.        _mint(caller, tokenId, quantity)
5.        NFBT[tokenId]. deviceId = deviceId
6.        NFBT[tokenId]. duration = duration
7.        NFBT[tokenId]. location = location
8.        NFBT[tokenId]. price = price
9.        tokenId += 1
```

Algorithm 1 presents the pseudocode for minting NFBTs. Within this contract, the Bluetooth service operator generates NFBTs with identical IDs by inputting parameters such as the service area address of the Bluetooth service device, device ID, price, duration, and the quantity to be minted.

**Algorithm 2:** buyNFBT

```
Input: tokenId, deviceId, amount , quantity
1.    if(deviceId = NFBT[tokenId]. deviceId) then
2.        if(amount = NFBT[tokenId]. price * quantity) then
3.            payable(owner).transfer(amount)
4.            _safeBatchTransferFrom(owner, caller, tokenId, quantity)
5.        if(now > User[tokenId]. expirationTime[caller]) then
6.            User[tokenId]. expirationTime[caller]) = now +
              NFBT[tokenId]. duration * quantity
7.        else if(now <= User[tokenId]. expirationTime[caller]) then
8.            User[tokenId]. expirationTime[caller]) += NFBT[tokenId].
              duration * quantity
```

Users can purchase NFBTs in the NFBT Client using ETH. Upon confirmation of the purchase, the request is transmitted to the blockchain network via the MetaMask wallet, and the smart contract verifies the payment amount, executes the asset transfers, and dynamically updates the user's access expiration time. Algorithm 2 presents the pseudocode for purchasing NFBTs.

**Algorithm 3:** softPairingEstablished

```
Input: NFBTId, deviceId
1.    require(caller owns the NFBT && the NFBT has not expired)
2.    require(the NFBT has not yet established an NFT Soft Pairing with
      the device && the device is active)
3.    linkedDeviceID[NFBTId] = deviceId
4.    deviceNFTContract.recordAccessLink(NFBTId, deviceId)
```

Algorithm 3 introduces an NFT soft pairing mechanism to establish an on-chain bidirectional mapping between NFBTs and NFDTs at the logical layer, thereby ensuring the uniqueness of the binding between NFBTs and NFDTs.

Fig. 3 shows an example of NFT soft pairing establishment and confirmation results, where NFBT ID is 1 and NFDT Device ID is Device1, and they have established an NFT soft pairing.

```
"from": "0x0f54baa70f2bf6cc6c438d389dfa9721eea3e4e6",
"topic": "0xa6175614e7f2e2d9e0e54eae70577c7381fe83e6ff423bad4a865b24b719cf70",
"event": "softPairingConfirmationRecorded",
"args": {
        "0": "Device1",
        "1": "1",
        "deviceId": "Device1",
        "NFBTId": "1"
}


"from": "0x756070820832d295c3c008dc0fe9d8a03ccf8895",
"topic": "0x7e75df6887da674da0eeccc37ca8013d80607e35e174b7419479dd822bfdcbdf",
"event": "softPairingEstablishedRecorded",
"args": {
        "0": "1",
        "1": {
                "_isIndexed": true,
                "hash": "0x9497a42c9cf17decdea11ed3bc80f51f2a5ba2f9f3c56df08a6a226cf814832a"
        }
}
```

Fig. 3. Soft pairing establishment and confirmation results.

**Algorithm 4:** accessVerification

```
Input: signAddress, tokenId, deviceId
1.    if(deviceId = NFBT[tokenId]. deviceId) then
2.        if(linkedDeviceID[tokenId] = deviceId) then
3.            if(balanceOf(signAddress, tokenId) >= 1) then
4.                if(now < User[tokenId].expirationTime[signAddress]) then
5.                    Access Verification Passed
```

The device performs an on chain query to verify whether a bidirectional NFT soft pairing relationship has been established and whether the signer address holds a valid NFBT. Upon successful verification, a Bluetooth connection is established. Algorithm 4 presents the pseudocode for NFBT verification.

### *B. NFDT Contract*

**Algorithm 5:** registerDevice

```
Input: deviceId, location, operator
1.    global tokenId
2.    require(device is unregistered && operator address != address(0))
3.    if(isAuthorized(caller)) then
4.        _mint(operator, tokenId, 1)
5.        deviceMetadata[tokenId]. deviceId = deviceId
6.        deviceMetadata[tokenId]. registeredTime = now
7.        deviceMetadata[tokenId]. location = location
8.        deviceMetadata[tokenId]. isActive = true
9.        tokenId += 1
```

Algorithm 5 is triggered by the Bluetooth service operator and initializes decentralized identity credentials for Bluetooth devices by binding the physical device identifier with the operator address on-chain.

**Algorithm 6:** softPairingConfirmation

```
Input: NFBTId, deviceId
1.    require(caller is NFBTContract && device is registered && device
      is active)
2.    require(NFBTContract.getDeviceId[NFBTId]=deviceId)
3.    deviceToNFBTs[deviceId].push(NFBTId)
```

Algorithm 6 performs cross-contract calls to verify the correspondence between access credentials and the target device, establishing an NFT soft pairing record with the NFBT on the device side to ensure the uniqueness and security of access authorization.

## IV. Security Analysis

This section evaluates the proposed framework against critical security threats:

1) Resistance to Unauthorized Service Access: Unlike conventional protocol-layer mitigation approaches, the proposed architecture performs authentication after device pairing but prior to connection establishment. Therefore, even if an attacker exploits inherent vulnerabilities to compromise the Bluetooth pairing process, access to Bluetooth services remains unattainable without both a valid cryptographic signature generated through the MetaMask wallet and verified ownership of a valid NFBT that has been soft paired with the target device. The relevant access control verification scenarios are presented in Table I.

TABLE I. Access control validation scenarios

| User Profile | Possesses NFBT | Token Status | Expected Result | Actual Result |
|---|---|---|---|---|
| Authorized | Yes | Active | Granted | Granted |
| Expired | Yes | Expired | Denied | Denied |

| Unauthorized | No | N/A | Denied | Denied |
|---|---|---|---|---|

2) Resistance to Replay Attacks and Centralization Risks: By transferring the trust anchor to the blockchain, the inherent limitations of centralized architectures are eliminated. Furthermore, the challenge–response protocol employs a dynamically generated random nonce, ensuring that intercepted historical signatures cannot be reused in subsequent authentication processes, thereby effectively mitigating replay attacks.

3) Dynamic Permission Revocation: The NFBT metadata supports time-based access control. Once the expiration time is reached, the NFT soft pairing relationship and the associated access privileges are atomically revoked according to the immutable on chain state, thereby enabling fine grained lifecycle management.

## V. Experimental Results And Analysis

### A. *Experimental Procedure*

To evaluate the feasibility and performance of the proposed NFBT-based Bluetooth access control system, a DApp named NFBT Client was developed. The smart contracts were deployed to a local Ethereum test network constructed using Ganache via the Remix development platform, with the Ethereum London hard fork rules enabled. The frontend of the NFBT Client is developed based on the Vue.js framework and integrated with the Web3.js library to enable interaction with the blockchain network. Since conventional web browsers cannot directly access blockchain networks, MetaMask was introduced as a browser-extension cryptocurrency wallet in the experiments. It is used to manage user accounts and private keys, and to provide functionalities such as transaction confirmation, random challenge signing, and account authorization. This enables secure interaction between the NFBT Client and the blockchain network while effectively preventing the exposure of users' private keys. On the Bluetooth device side, a Raspberry Pi 4B is utilized as the Bluetooth service device, which hosts the proposed authentication backend for challenge generation, signature verification, blockchain state querying, and Bluetooth connection management. Furthermore, to ensure that all Bluetooth connections are controlled through the designed authentication procedure, the system restricts users from establishing Bluetooth connections directly via system settings. Consequently, users are required to initiate access requests exclusively through the developed NFBT Client, thereby ensuring that the entire access process strictly adheres to the proposed authentication mechanism.

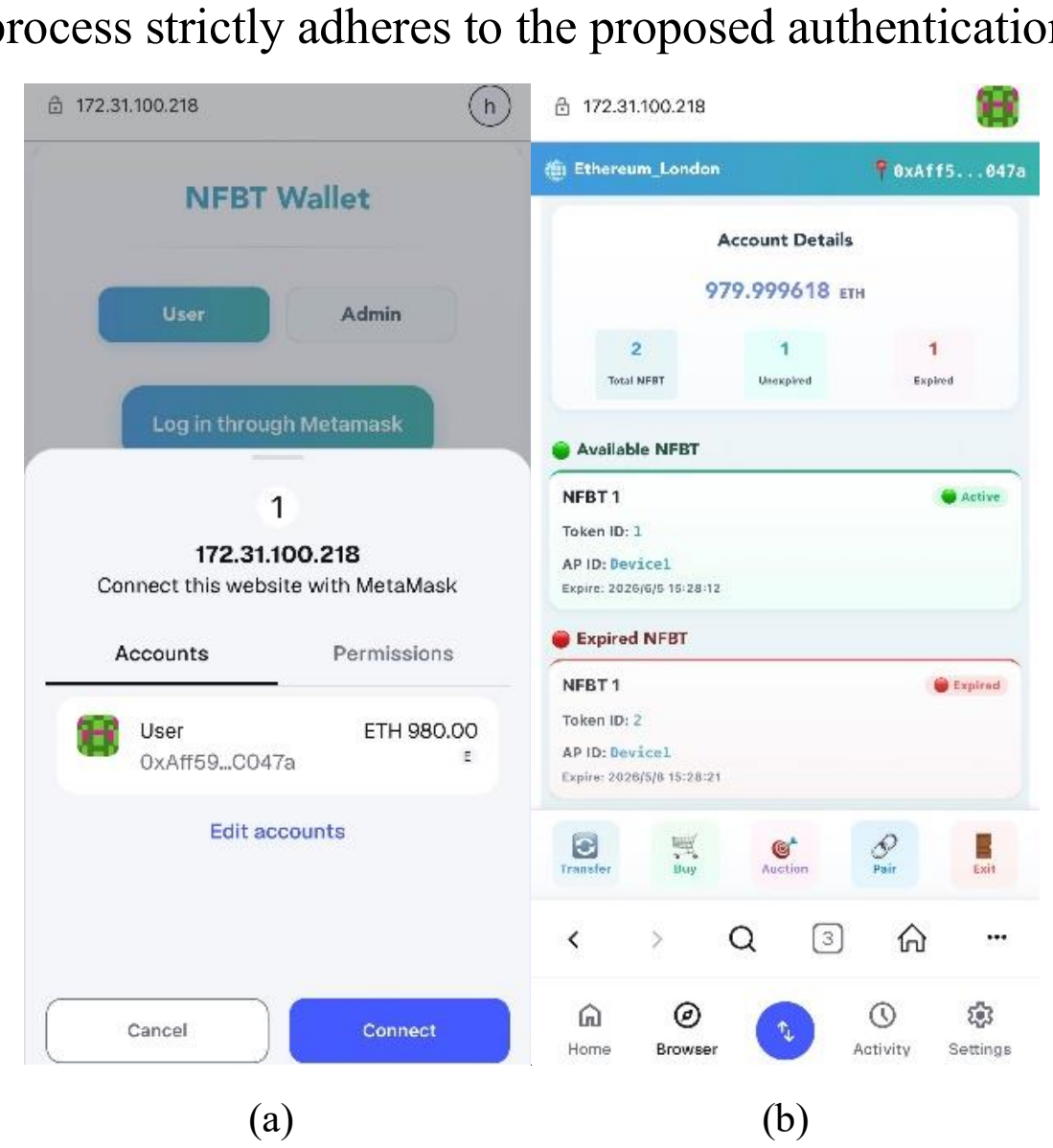

(a) (b)

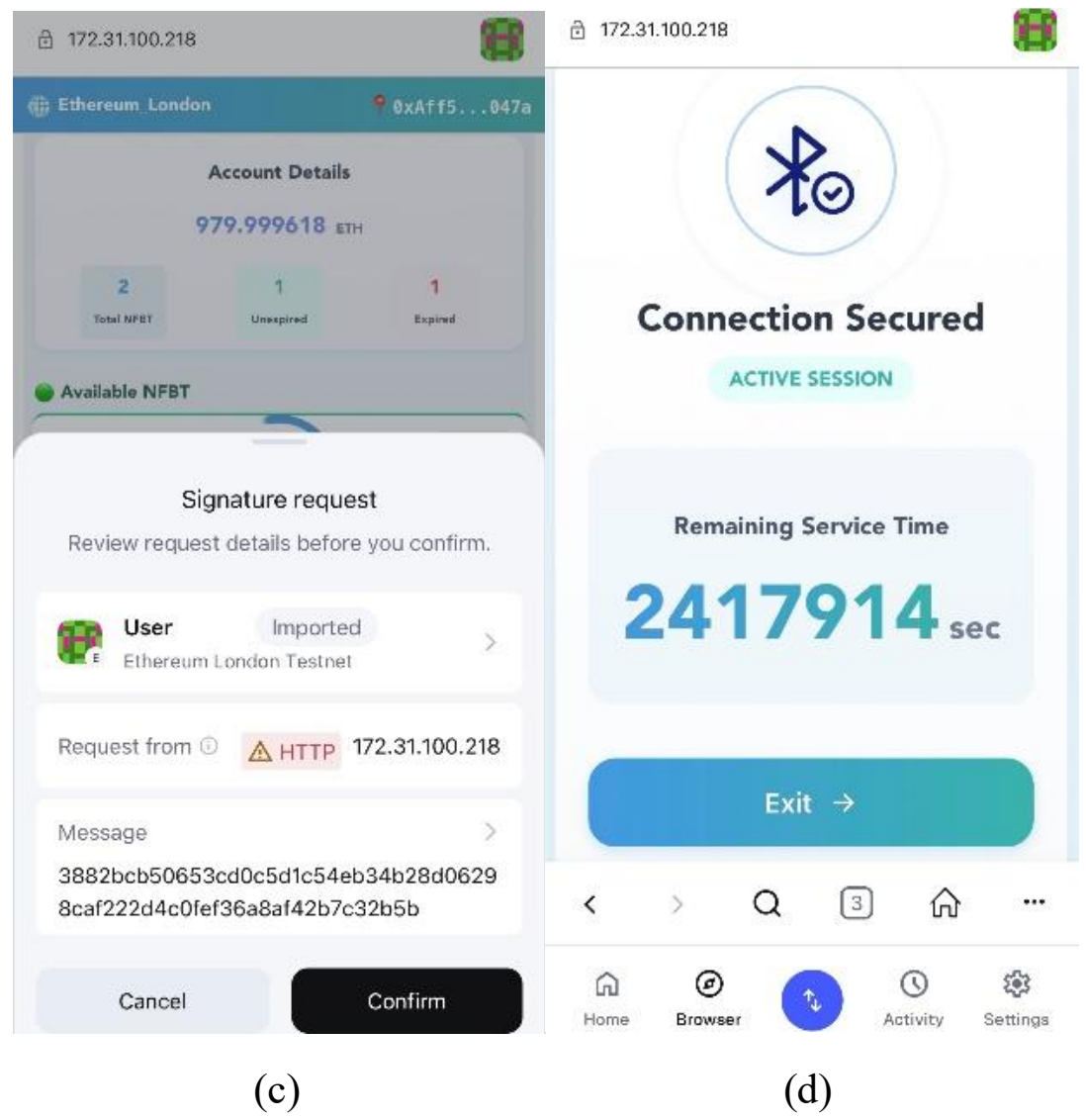

(c) (d)

Fig. 4. Experimental procedure. (a) User login. (b) User Information Page. (c) Sign the challenge value. (d) Connection success page.

As illustrated in Fig. 4(a), the user with address 0xAff5…047a can log into the NFBT Client via MetaMask. Fig. 4(b) presents an example interface of the user's Bluetooth wallet client, which lists all NFBTs currently owned by the user along with their corresponding status information, and integrates relevant management and operational functionalities. Among them, the NFBT with ID 2 has expired and is no longer valid for Bluetooth access verification, whereas the token with ID 1 remains valid and can be used to establish a connection with the Bluetooth service device identified by Device1.

When the user selects the NFBT with ID 1 to request access to the target Bluetooth service device, the user device, the Bluetooth service device, and the blockchain network perform access permission verification according to the interaction procedure described in Fig. 2. Fig. 4(c) illustrates the user signing the challenge value via MetaMask. Upon successful verification, as shown in Fig. 4(d), the Bluetooth service device returns the verification result to the user and displays authorization information, such as the remaining available connection time, in the Bluetooth wallet client. The user can then proceed to utilize the Bluetooth services provided by the device.

## B. Performance Analysis

The authentication delay, denoted as $T_{delay}$, is defined as the time interval from when a user initiates a Bluetooth access request to when the Bluetooth device successfully completes the verification process. To accurately evaluate system performance, 20 repeated experiments were conducted in a controlled network environment, and the average time consumption of each stage was recorded and analyzed.

The total delay of a single authentication process, $T_{delay}$, can be expressed as follows:

$$T_{delay} = T_{challenge} + T_{sign} + T_{verify}$$

where $T_{challenge}$ represents the time required for the Bluetooth device to generate the random challenge value, $T_{sign}$ denotes the time for the user to complete signature authorization via MetaMask, and $T_{verify}$ corresponds to the total time consumed by the Bluetooth device for address recovery, on-chain NFBT state querying, and access permission verification. The measured average overhead of each stage is presented in Table II.

TABLE II. AUTHENTICATION DELAY OVERHEAD

| $T_{challenge}$ | $T_{sign}$ | $T_{verify}$ |
|---|---|---|
| 0.335ms | 990.26ms | 104.36ms |

The experimental results indicate that the overall access latency of the system is primarily constrained by the user-side signing phase, which is mainly attributed to the manual confirmation process in MetaMask interactions and the computational overhead of signature generation. In contrast, the verification phase based on blockchain state querying employs read-only view calls, which do not require state modification or block confirmation. Therefore, it exhibits low and relatively stable latency, approximately 104ms, demonstrating the efficiency of the lightweight on-chain query mechanism. Although the proposed scheme introduces an additional latency of approximately 1.09s compared with conventional Bluetooth direct connections without authentication, this overhead fundamentally represents a trade-off between access efficiency and security.

In the experiments, the Bluetooth service device periodically checked the authorization status of devices in the control set with a polling interval of 1s. When an NFBT expired, the device automatically terminated the Bluetooth connection, thereby enabling a dynamic permission revocation mechanism without manual intervention.

By sacrificing a marginal amount of initial connection efficiency, the system achieves decentralized strong identity authentication. Even when the pairing mechanism is compromised, the proposed scheme can effectively prevent unauthorized malicious connections while simultaneously enabling fine-grained permission revocation, thereby enhancing the security of Bluetooth access.

TABLE III. GAS CONSUMPTION OF THE MAIN FUNCTIONS

| Function | Gas Consumption |
|---|---|
| mintNFBT() | 134625 gas |
| buyNFBT() | 95598 gas |
| softPairingEstablished() | 132615 gas |
| registerDevice() | 197441 gas |

In the Ethereum blockchain, the execution cost of each transaction is determined by transaction complexity, data payload, and the computational cost of function execution. Table III presents the gas consumption of the core functionalities in the proposed system, including NFBT minting, purchasing, bidirectional NFT soft pairing establishment and device registration. The experimental results indicate that the registerDevice function incurs the highest cost, which is primarily attributed to the on-chain persistent storage of critical state variables, such as token metadata and device association information. Since each function embodies specific application logic, such as resource acquisition, access control, and traceability verification, gas consumption is not directly comparable across functions, and its cost exhibits strong function-dependent characteristics.

The experimental results demonstrate that the gas consumption of ERC721 increases linearly with the number of tokens minted, from 152,960 gas to 2,055,000 gas, whereas ERC1155 exhibits a constant cost characteristic when minting credentials with the same ID, remaining stable at 134,625 gas. When minting 15 credentials in batch, the proposed scheme reduces gas consumption by approximately 93.4% compared to the conventional approach. This high level of cost efficiency effectively offsets the additional overhead introduced by enhanced security. Such low-cost and high-efficiency distribution capability enables the system to achieve stronger security protection than traditional centralized architectures without introducing additional operational burdens.

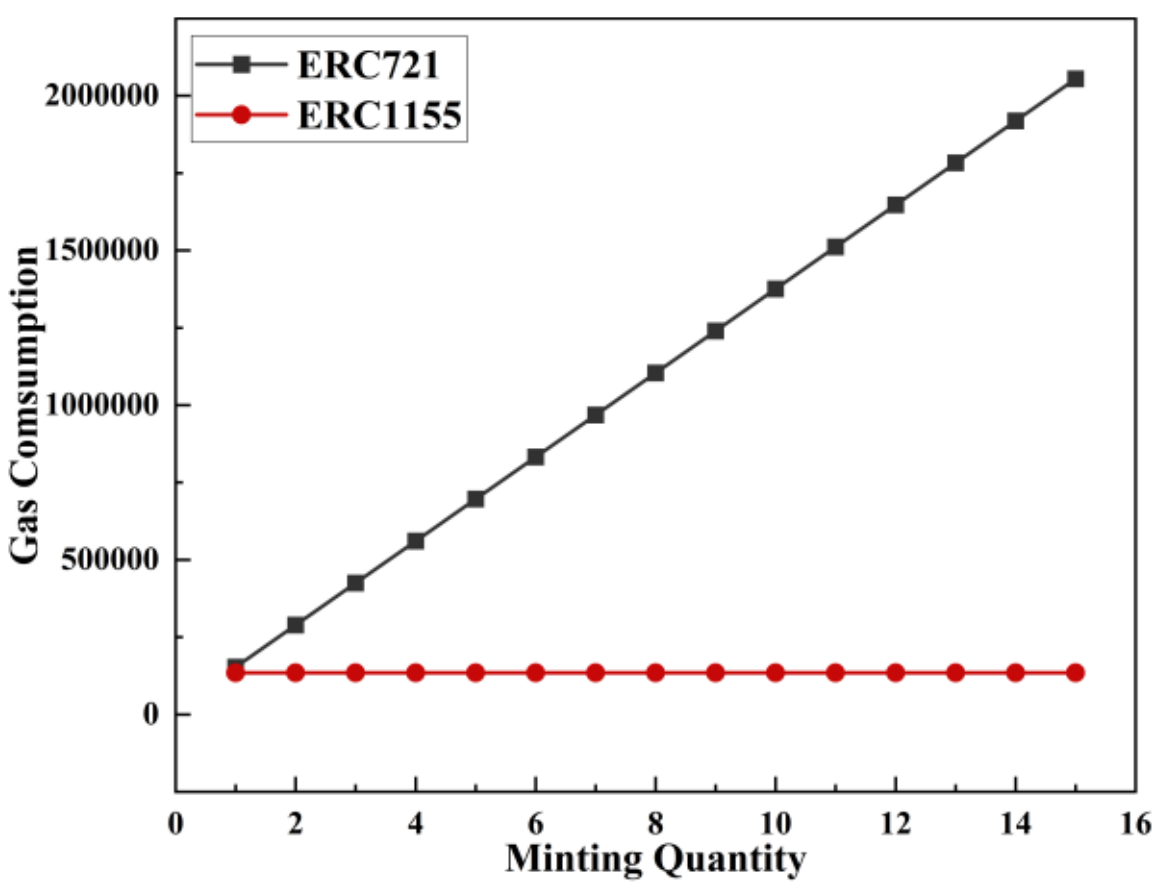

Fig. 5. Gas consumption of tokens minted according to different standards.

## VI. CONCLUSION

This paper proposes an NFT soft pairing framework for Bluetooth access control that decouples native Bluetooth pairing from service authorization through a three-layer architecture. In this framework, the Bluetooth layer preserves native pairing and communication, the blockchain layer provides trusted execution and on-chain state verification, and the application layer implements NFT soft pairing to define authorization logic. NFBTs and NFDTs form a bidirectional relationship between user credentials and device identities, where access is no longer implicitly granted by pairing status but explicitly governed by on-chain NFT soft pairing state and cryptographic verification. The proposed design preserves compatibility with existing Bluetooth infrastructure while enabling dynamic, revocable, and fine-grained access control. A prototype implementation demonstrates secure authentication, efficient revocation, and acceptable performance overhead. Overall, NFT soft pairing provides a general abstraction that separates connectivity from trust in Bluetooth systems.

## ACKNOWLEDGMENT

Guangdong Province Graduate Education Innovation Program Project[2024JGXM_163]. Shenzhen University High-Level University Construction Phase III -Human and Social Sciences Team Project for Enhancing Youth Innovation[24QNCG06].